\documentclass[aps,prl,twocolumn,showpacs,superscriptaddress,sortedaddress]{revtex4-1}
\pdfoutput=1
\usepackage{graphicx}  			
\usepackage{dcolumn}   			
\usepackage{amssymb,amsmath,bm}	
\usepackage{epstopdf}  			
\usepackage{color}
\usepackage{times}
\begin{document}

\title{Bond disorder induced criticality of the three-color Ashkin-Teller model}

\author{Arash Bellafard} 
\affiliation{Department of Physics and Astronomy, University of California, Los Angeles, CA 90095, USA}

\author{Helmut G.~Katzgraber}
\affiliation {Department of Physics and Astronomy, Texas A\&M University,
College Station, Texas 77843-4242, USA}
\affiliation {Theoretische Physik, ETH Zurich, CH-8093 Zurich, Switzerland}

\author{Matthias Troyer}
\affiliation {Theoretische Physik, ETH Zurich, CH-8093 Zurich, Switzerland}

\author{Sudip Chakravarty}
\affiliation{Department of Physics and Astronomy, University of California, Los Angeles, CA 90095, USA}

\date{\today}

\begin{abstract}
An intriguing  result of statistical mechanics is that a first-order phase transition can be rounded by disorder coupled to energy-like variables. In fact, even more intriguing  is that  the rounding may manifest itself as a critical point, quantum or classical. In general,  it is not known, however, what universality classes, if any, such criticalities belong to. In order to shed light on this question we examine in detail the disordered three-color Ashkin-Teller model by Monte Carlo methods. Extensive analyses indicate that the critical exponents define a new universality class.	
We show that the rounding of the first-order transition of the pure model due to the impurities is manifested as criticality. However, the magnetization critical exponent, \(\beta\), and the correlation length critical exponent, \(\nu\),  are found to vary with disorder and the four-spin coupling strength, and we  conclusively rule out that the model belongs to the universality class of the two-dimensional Ising model. 
\end{abstract}


\maketitle

Rounding of first-order transitions due to disorder is an important problem because such transitions are  ubiquitous in both classical and quantum systems. In contrast to the effect of disorder on continuous transitions, much less is known about its effect on first-order  transitions.  Imry and Wortis~\cite{imry:1979}, Hui and Berker~\cite{Hui:1989}, and Aizenman and Wehr~\cite{Aizenman:1989,*Aizenman:1990} have made important contributions to this subject. 
 
More recently, the quantum  $N$-color Ashkin-Teller (AT) model in one dimension was studied using both weak and strong disorder analyses~\cite{Goswami:2008} where evidence was presented in favor of rounding of this  transition into a quantum critical point for $N\ge 3$. An important  question that is unanswered is   the universality class of this transition. It was argued, however, that the physical picture is the same as the random transverse-field Ising model for a limited parameter regime. Even more recently, general and rigorous analyses have shown that disorder can indeed {\em round} a first-order quantum phase transition~\cite{Greenblatt:2009,*Aizenman:2012}. However, it is still unknown what the possible universality classes are, or if rounding is not simply a smearing of the first-order transition instead of an emergent criticality. 
 
The phase transition in the \emph{pure} $N$-color AT model has been studied for decades~\cite{Grest:1981,*Fradkin:1984,*Shankar:1985}. It is known that for $N \ge 3$, the pure model undergoes a first-order phase transition in two dimensions ($2D$). However, the  corresponding disordered case is still far from being understood.  Here, we address the criticality of the bond disordered classical three-color AT model in $2D$, a necessary prerequisite for understanding the corresponding quantum phase transition at zero temperature. 
Utilizing the Monte Carlo method, we compute the critical exponents for the magnetization and the localization length, and show that the phase transition does not belong to the Ising universality class, as previously suggested~\cite{Cardy:1996,*Cardy:1999}. We also find that the critical exponents vary with disorder and the four spin coupling strength. Previously  a perturbative  two-loop  renormalization group calculation had  hinted at a new strong-disorder fixed point~\cite{Pujol:1996}. That the Ising universality class does not hold  appears to be
similar to the $q$-state Potts model~\cite{Olson:1999} in the presence of disorder, where the exponent also  varies with $q$.

The  Hamiltonian of the $N$-color AT model in $2D$ is 
\begin{equation}\label{eq:hamiltonian}
\mathcal{H} = -\sum_\alpha \sum_{\langle i,j \rangle} J_{i,j} s_i^\alpha s_j^{\alpha} - g \sum_{\alpha\neq\beta} \sum_{\langle i,j\rangle} s_i^\alpha s_j^\alpha s_i^\beta s_j^\beta ,
\end{equation}
where the classical spin variables $ s_i^\alpha = \pm 1 $ reside on a square lattice, and the  \( N \) spins on a single lattice site are labeled by  $ \alpha, \beta = 1, 2, \dots, N $. The sum is only over nearest neighbor pairs. In our study, $N=3$ and we only consider the case for $ g > 0 $. The ferromagnetic coupling constant $ J_{i,j} > 0 $ is a random variable following a translation-invariant binary probability distribution  of the form:
\begin{equation}
	p[J_{i,j}] = \begin{cases}
		J-\frac{\Delta}{2}, &\text{ with probability } 1/2\\
		J+\frac{\Delta}{2}, &\text{ with probability } 1/2.
	\end{cases}
\end{equation}

The magnetization of the system is
\begin{equation}
	m = \frac{1}{N}\left[ \left\langle \sum_{\alpha=1}^N \left| m_\alpha \right| \right\rangle \right],
\end{equation}
where the symmetry between spins of different colors is utilized to increase accuracy. The angular brackets \( \left\langle \cdots \right\rangle \) denote the usual thermal Monte Carlo average, whereas the square brackets \( \left[ \cdots \right] \) denote the quenched average over configurations with different \( \{ J_{i,j} \} \).

Close to a continuous transition, the magnetization \( m \) scales as
\begin{equation}\label{eq:m_scaling_form}
	m = L^{-\beta/\nu}\widetilde m \left( x L^{1/\nu} \right)
\end{equation}
where \( \widetilde m(\cdot) \) is a universal function, and \( \nu \) is the critical exponent for the correlation length. The variable   \( x = ( J-J_c ) / J_c \) is the reduced coupling constant in temperature units.

There are two other quantities of interest that are useful in determining the nature of the phase transitions: the energy cumulant \( V_\epsilon \) and magnetic cumulant \( V_m \)~\cite{Binder:1984,Challa:1986} given by 
\begin{equation}\label{eq:binder_ratio}
	V_\epsilon = 1-\frac{\left[ \left\langle \epsilon^4 \right\rangle \right]}{3\left[ \left\langle \epsilon^2 \right\rangle \right]^2}, \qquad V_m = 1-\frac{\left[ \left\langle m^4 \right\rangle \right]}{3\left[ \left\langle m^2 \right\rangle \right]^2},
\end{equation}
respectively, where \( \epsilon = \mathcal{H}/(NL^2) \), \( \mathcal{H} \) is the Hamiltonian given by Eq.~\eqref{eq:hamiltonian}, and $L$ is the linear dimension of the square lattice.
For a continuous transition, the cumulant $V_m$ has the scaling form
\begin{equation}\label{eq:gm_scaling_form}
	V_m = \tilde V_m \left( x L^{1/\nu} \right).
\end{equation}
Using $V_\epsilon$ we can determine the order of the phase transition. In a continuous phase transition, \( V_\epsilon \rightarrow 2/3 ~\forall~ J \) as \( L \rightarrow \infty \), whereas in a first-order phase transition, \( V_\epsilon \) behaves  like the continuous case \emph{except} at \( J = J_c \), where \( V_\epsilon \) approaches a non-universal constant. 

To apply the cluster Monte Carlo algorithm to our system, we fix a color \(A\) and rewrite the Hamiltonian in Eq.~\eqref{eq:hamiltonian} to obtain
\begin{eqnarray}\label{eq:modified_hamiltonian}
	\mathcal H &=& \mathcal H_{\bar A} - \sum_{\langle i,j \rangle} \left( J_{i,j} + g\sum_{\alpha\neq A} s_i^\alpha s_j^\alpha \right) s_i^A s_j^A\\
	                    &=& \mathcal H_{\bar A} + \sum_{\langle i,j \rangle}E_{i,j}^A,
\end{eqnarray}
where \( H_{\bar A} \) represents the terms which do not include spins of color \( A \), and $E_{i,j}^A$ is the bond energy for the bond between sites \( i \) and \( j \) in color \( A \). This expression can be viewed as a random Ising model Hamiltonian with equivalent nearest-neighbor coupling constant \( J_{i,j} + g\sum_{\alpha\neq A} s_i^\alpha s_j^\alpha \) if only color \( A \) is considered. Therefore, any cluster algorithms for the random Ising model can be adopted here. For the following calculation, the algorithm proposed by Niedermayer~\cite{Niedermayer:1988} is used. Note that the bond energy is bounded from above by
\begin{equation}
	E_\text{max} = J + \frac \Delta 2 + (N-1)g.
\end{equation}

The equilibration time for all observables are found by logarithmic binning. We perform one million Monte Carlo steps for equilibration. Furthermore, we perform \(10,000\) thermal averages over each disordered configuration. The number of disorder averages used for each observable varies and is given in the caption of the plots.

Because all temperatures are simulated with the same disorder realization, the measured data are correlated. Therefore, in order to obtain the error for the cumulants, we have applied the Jacknife procedure to correct for bias~\cite{Wu:1986}.

\begin{figure}[!htb]
	\centering
	\includegraphics[width=1\linewidth]{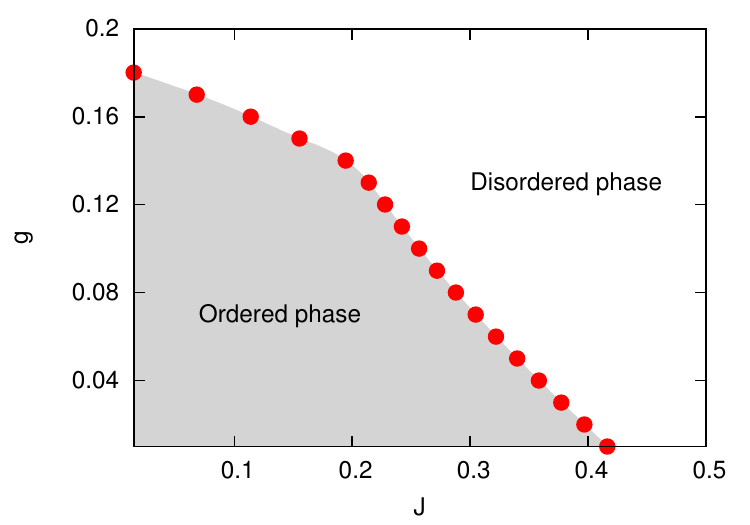}
	\caption{(Color online) Phase diagram of \( g \) versus \( J \) with \( \Delta = 0 \) for \(N=3\). For any values of $g \lesssim 0.18$, the system undergoes a first-order phase transition. The error bars are smaller than the size of the symbols used in the plot.}
	\label{fig:phSp}
\end{figure}

First, we must identify the parameter regime in which the pure system undergoes a first-order transition by looking at the phase space of the pure AT Hamiltonian. For fixed values of \(g\), we calculate \(V_\epsilon\) for system sizes \(L = 24, 32, \dots, 64\). The first-order transition is confirmed by inspecting how the depth of the energy cumulant, \( 2/3 - V^*_\epsilon[J_c(L),L] \), changes as a function of some inverse power of the system size. As expected, the extrapolated lines intersect the ordinate at some nonzero finite value. This validates that the transitions are all first-order for \( g \lesssim 0.18 \). Moreover, expecting the transition point to be close to the minimum of \(V_\epsilon\) for the largest examined size, we deduce the phase diagram of the pure AT Hamiltonian as shown in Fig.~\ref{fig:phSp}.

\begin{figure}[!htb]
	\centering
	\includegraphics[width=1\linewidth]{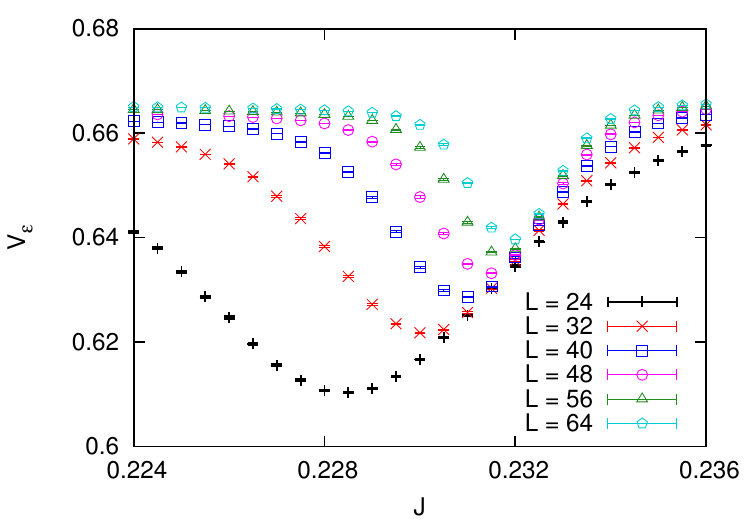}
	\caption{(Color online) The energy cumulant \( V_\epsilon \) versus \( J \) with \( (g,\Delta) = (0.1,0.2) \) for \(N=3\). The measured values are averaged over $15,000$ configurations. \(V_\epsilon\) approaches the value $2/3$ for all $J$ in the continuous phase transition case and eventually becomes $2/3$ for infinite system size. The error bars are hardly visible because they are smaller than the symbols.}
	\label{fig:g01_d02_ge}
\end{figure}

\begin{figure}[!htb]
	\centering
	\includegraphics[width=1\linewidth]{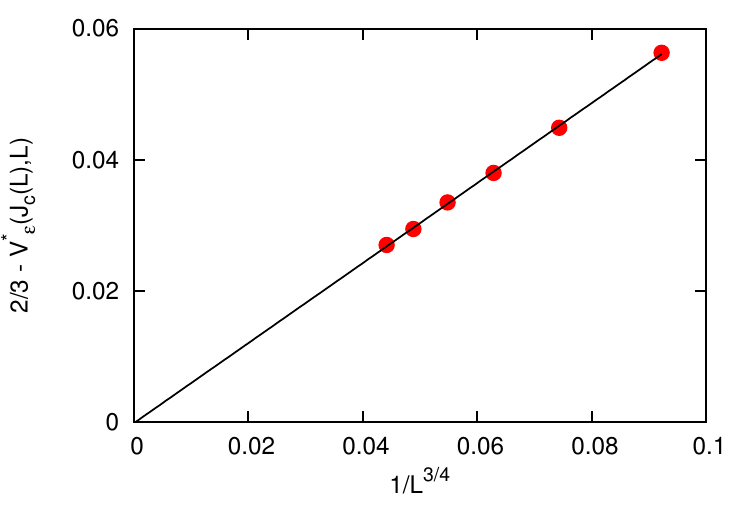}
	\caption{(Color online) \( (2/3 - V^*_\epsilon(J_c(L),L)) \) versus \( 1/L^{3/4} \) with \( (g,\Delta) = (0.1,0.2) \) for \(N=3\). The line goes through the origin indicating a second-order phase transition. The error bars are smaller than the size of the symbols used in the plot.}
	\label{fig:g01_d02_ge_star}
\end{figure}

Now we study the disordered system for the parameter set \( (g,\Delta) = (0.1,0.2) \) and use \( V_\epsilon \), shown in Fig.~\ref{fig:g01_d02_ge}, to find the order of the transition. Plotting the depth of \(V_\epsilon\) against \( 1/L^{3/4} \) (Fig.~\ref{fig:g01_d02_ge_star}), we observe that \( V^*_\epsilon \rightarrow 2/3 \) as \( L \rightarrow \infty \), which indicates that the transition is continuous.

\begin{figure}[!htb]
	\centering
	\includegraphics[width=1\linewidth]{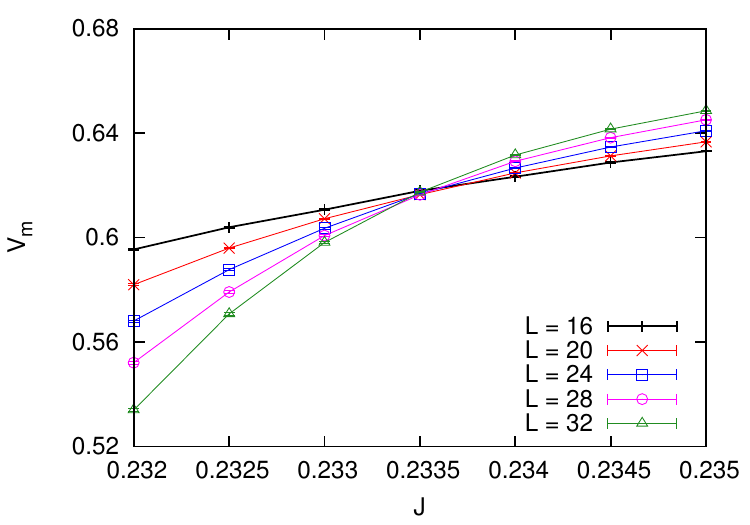}
	\caption{(Color online) Magnetic cumulant \( V_m \) versus \( J \) with \( (g,\Delta) = (0.1,0.2) \) for \(N=3\). The measured values are averaged over $49,000$ configurations. The intersection point is at  $J_c = 0.2334 \pm 0.0003$. The error bars are hardly visible because they are smaller than the symbols.}
	\label{fig:g01_d02_gm}
\end{figure}

\begin{figure}[!htb]
	\centering
	\includegraphics[width=.9\linewidth]{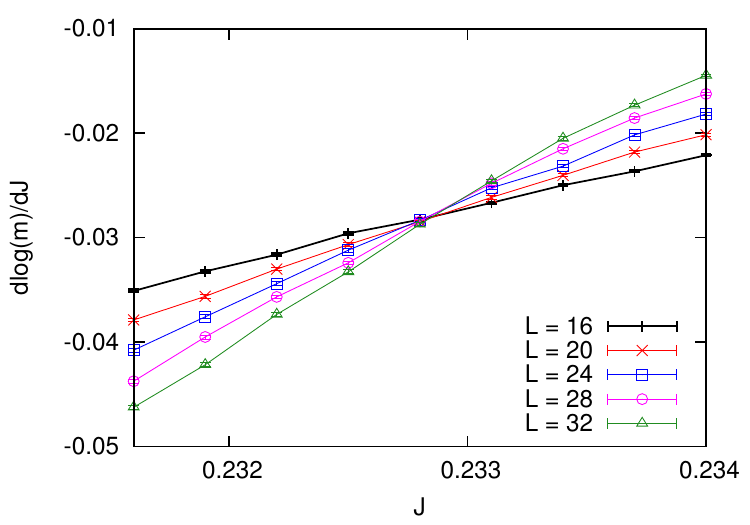}
	\caption{(Color online) Logarithmic derivative of magnetization \( d\log(m)/dJ \) versus \( J \) with \( (g,\Delta) = (0.1,0.2) \) for \(N=3\). The measured values are averaged over $20,000$ configurations. The intersection point is at \( J_c = 0.2328 \pm 0.0003 \). The error bars are hardly visible because they are smaller than the symbols.}
	\label{fig:g01_d02_d_log_m_d_j}
\end{figure}

We use the scaling properties of \(V_m\) in Eq.~\eqref{eq:gm_scaling_form}, and the logarithmic derivative of the magnetization squared to find an accurate estimate of the transition point \( J_c \). Figures ~\ref{fig:g01_d02_gm} and ~\ref{fig:g01_d02_d_log_m_d_j} show that the two quantities are independent of the system size at a single critical point, \( J_c = 0.2331 \pm 0.0006 \).

\begin{figure}[!htb]
	\centering
	\includegraphics[width=1\linewidth]{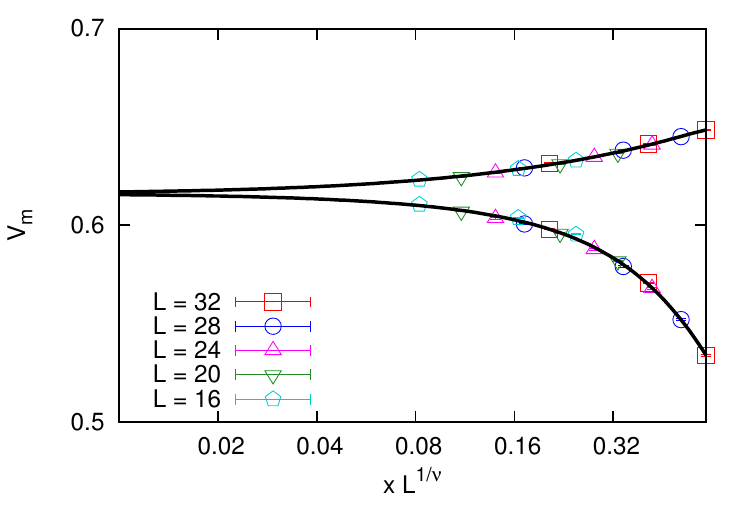}
	\caption{(Color online) Finite size scaling plot of the magnetic cumulant with \((g,\Delta) = (0.1, 0.2)\) for \(N=3\), \(J_c = 0.2335\), and \(\nu = 0.76\). The upper branch corresponds to the ordered phase whereas the lower branch corresponds to the disordered phase. The error bars are hardly visible because they are smaller than the symbols.}
	\label{fig:gm_fss}
\end{figure}

\begin{figure}[!htb]
	\centering
	\includegraphics[width=1\linewidth]{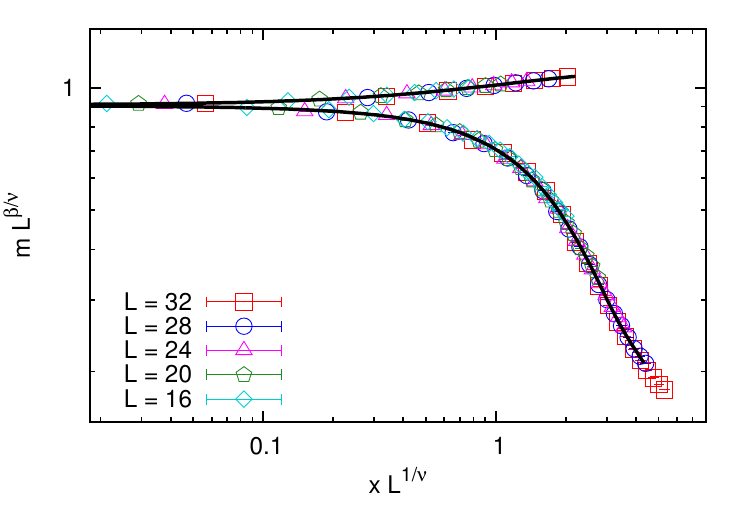}
	\caption{(Color online) Finite size scaling plot of the magnetization with \((g,\Delta) = (0.1, 0.2)\) for \(N=3\), \(J_c = 0.2334\), and \( \beta/\nu \approx 0.07 \). The magnetization values are averaged over $10,000$ configurations. The upper branch corresponds to the ordered phase whereas the lower branch corresponds to the disordered phase. The error bars are hardly visible because they are smaller than the symbols.}
	\label{fig:g01_d02_m_fss}
\end{figure}

From the scaling behavior of \( V_m \) in Eq.~\eqref{eq:gm_scaling_form}, we deduce the critical exponent \( \nu \). The best data collapse is obtained with $J_c = 0.2335 \pm 0.0001$ and $\nu = 0.76 \pm 0.05$, see Fig.~\ref{fig:gm_fss}. Furthermore, the critical exponents \( \beta \) and \( \nu \) can be extracted from Eq. \eqref{eq:m_scaling_form}. As shown in Fig.~\ref{fig:g01_d02_m_fss}, the magnetization scales well with \( \nu = 0.70 \pm 0.02 \) and \( \beta = 0.055 \pm 0.005 \). 

To address the universality of the exponents, it is important to study the behavior of the system for more than just one particular set of parameters \( (g, \Delta) \). In this regard, we repeat our calculation for the parameter sets \( (g,\Delta) = (0.05, 0.20) \) and \( (0.10, 0.10) \). Table~\ref{tbl:exponents} summarizes the values of \(\nu\) and \(\beta\) for each parameter set.

\begin{table}[h]
\centering
\begin{tabular}{l |l |l |l}
\hline\hline
$(g,\Delta)$	&	$(0.10, 0.20)$				&	$(0.05, 0.20)$				&	$(0.10, 0.10)$\\
\hline\hline
$V_m$			&	$\nu = 0.76 \pm 0.05$ 		&	$\nu = 0.71 \pm 0.01$ 		&	$\nu = 0.59 \pm 0.04$\\
\hline
$m$				&	$\nu = 0.70 \pm 0.02$		&	$\nu = 0.79 \pm 0.02$		&	$\nu = 0.53 \pm 0.01$\\
				&	$\beta =  0.055 \pm 0.005$	&	$\beta =  0.080 \pm 0.005$	&	$\beta = 0.060 \pm 0.007$\\
\hline\hline
\end{tabular}
\caption{The values of critical exponents extracted by finite-size scaling of magnetic cumulant and magnetization for different values of coupling constants and disorder.}
\label{tbl:exponents}
\end{table}

Based on our results, the disordered three-color AT model does not belong to the Ising universality class. One is tempted to conclude that reducing the coupling constant \(g\) has no effect on the exponent \(\nu\) (at least within the error bars) but results in an increase in the exponent \(\beta\). On the other hand, reducing the disorder \(\Delta\) has a lesser impact on the exponent \(\beta\), but it results in smaller \(\nu\). 

\begin{figure}[!htb]
	\centering
	\includegraphics[width=1\linewidth]{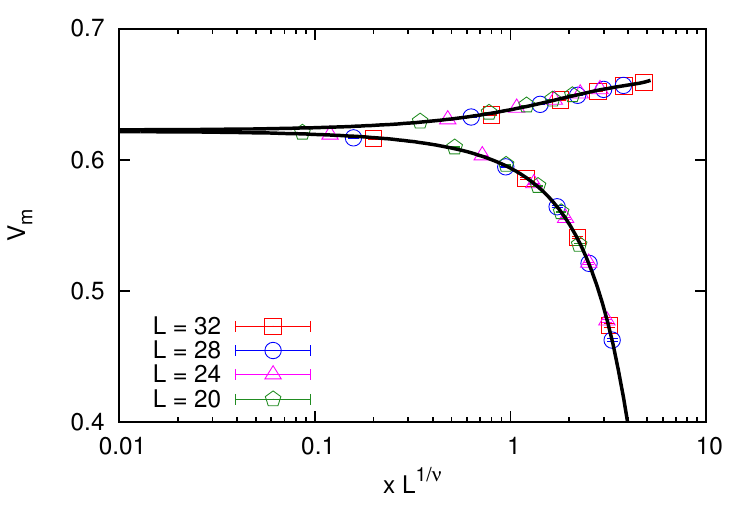}
	\caption{(Color online) Finite size scaling plot of the magnetic cumulant with \((g,\Delta) = (0.1, 0.1)\) for \(N=3\), \(J_c = 0.2436\), and \(\nu = 0.56\). The upper branch corresponds to the ordered phase whereas the lower branch corresponds to the disordered phase. The error bars are hardly visible because they are smaller than the symbols.}
	\label{fig:gm_fss_2}
\end{figure}

The results, when taken at face value, seem to violate the bound \( \nu \ge 2/D \)~\cite{Chayes:1986}. The value of $\nu$ is extremely sensitive to $J_c$: it is remarkable that a variation of $J_{c}$ from $0.2334$ to $0.2336$ can change $\nu$ from $0.71$ to $0.81$. In contrast, the same variation of $J_c$ results in a much weaker variation of the exponent $\beta$ from Eq.~\eqref{eq:m_scaling_form}. The spread of $\beta$ is in the range of $0.05$ to $0.06$ which is considerably smaller than the Ising exponent $1/8$.

Another striking fact is that the critical exponents \(\beta\) and \(\nu\) vary as the coupling parameter \(g\) or the disordered strength \(\Delta\) is altered. Similar behavior has been previously reported in studies of the critical behavior of the $q$-states Potts model~\cite{Cardy:1997,*Jacobsen:1998}, where the magnetic exponent varies continuously with \(q > 4\). Interestingly, in the very same studies the bound \( \nu \ge 2/D \) was violated, as in our present analysis.

It is worth mentioning that we have also examined the finite-size scaling of the correlation length $\xi$ extracted from the second moment correlation function~\cite{Ballesteros:00}
\begin{equation}\label{eq:structure}
	S(k) = \frac{1}{L^2} \sum_{i,j} e^{i\bf{k} \cdot (\bf{R}_i - \bf{R}_j)} \langle S_i S_j \rangle^2.
\end{equation}
The correlation length scales as $\xi/L= \tilde{\xi}\left(xL^{1/\nu}\right)$ for a continuous transition. Although preliminary results for $\nu$ are similar, the larger system sizes are much harder to achieve with our limited computation time, since the sum in Eq.~\eqref{eq:structure} is to be performed over all lattice points. The multifractality of the correlation function~\cite{Olson:1999} may be another difficulty in extracting reliable results. 

In summary, although precise values of the exponents are difficult to establish, the emergence of criticality induced by disorder in the pure AT model for \(N = 3\) (and presumably for \(N>3\) as well) is manifest and the presence of a single universality class (such as Ising) appears to be excluded. Further analytical and numerical studies of the corresponding quantum version~\cite{Goswami:2008} remain to be conducted.

A.B. and S.C. thank the National Science Foundation, Grant No.~DMR-1004520 for support. H.G.K.~acknowledges support from the Swiss National Science Foundation (Grant No.~PP002-114713) and the National Science Foundation (Grant No.~DMR-1151387). We thank ETH Zurich and UC Los Angeles for CPU time on the Brutus and Hoffman2 clusters. M.T. and S.C. also thank the Aspen center for physics for its hospitality.

%

\end{document}